\documentclass[preprint,floatfix]{revtex4}
\usepackage{graphicx}
\usepackage{epsf}
\newcommand{\dt}{{\Delta t}}

\newcommand{\bea}{\begin{eqnarray}}
\newcommand{\eea}{\end{eqnarray}}
\newcommand{\be}{\begin{equation}}
\newcommand{\ee}{\end{equation}}
\newcommand{\ba}{\begin{eqnarray}}
\newcommand{\ea}{\end{eqnarray}}

\newcommand{\nn}{\nonumber}
\newcommand{\la}{\label}

\def\t1{e_{_T}}
\def\v1{e_{_V}}

\begin{document}
\title{Elementary derivations of centripetal acceleration and\\ works of Huygens and Newton:
 a revisit with \\ a simple physical derivation}

\author{Siu A. Chin}

\affiliation{Department of Physics and Astronomy,
Texas A\&M University, College Station, TX 77843, USA}

\begin{abstract}

This work shows how two simple derivations of the centripetal acceleration are 
fundamentally related to seminal works of Huygens and Newton. 
A different, more physically motivated derivation is then given.  

\end{abstract}
\maketitle


The continued interest in elementary (no calculus, no vector, no trig function) 
derivations\cite{jon69,pat75,hab81,tip91,wed93,bro94,kra95,hen00,lef02,cor12} 
of the centripetal acceleration, seems not just an effort to teach a difficult concept to beginning physics students,
but actually reflects how circular motion was originally understood by founders\cite{wes71,erl94,sti01} 
of classical mechanics. It seems pedagogically enriching that this historical 
connection should be made more widely known among students and teachers of centripetal acceleration,
especially when these elementary derivations are used. 

The works of Huygens and Newton on centripetal acceleration are of course
well known to historians.\cite{wes71,erl94,sti01} However, they are not aware or interested 
in modern derivations. For beginning students of physics, Ref.\onlinecite{hen00} has given an excellent 
historical introduction on centripetal acceleration. This work goes beyond Ref.\onlinecite{hen00}'s discussion 
by providing precise details on Newton's\cite{wes71,erl94,sti01} and Huygens'\cite{sti01,erl94,huy} contributions.   

The simplest derivation of the centripetal acceleration
\be
a=\frac{v^2}{r}
\la{acc}
\ee
is undoubtedly the two circle method,\cite{wed93} which has been
independently derived numerous times.\cite{hab81,bro94,kra95,hen00}
The above acceleration follows when the angular velocity $2\pi/T$ in the circle traced out 
by the position vector
\be
v=\frac{2\pi r}{T}
\la{ac}
\ee
is substituted into the circle traced out by velocity vector
\be
a=\frac{2\pi v}{T}.
\la{aa}
\ee
However, despite its simplicity, (\ref{aa}) is usually not invoked in textbooks
because a circle in ``velocity space''\cite{wed93} was deemed too abstract for most students. 
Below we will first show that Newton already knew about the velocity circle (\ref{aa}) from his 
earliest study of circular motion.  
 
Since constant angular velocity is the key,
acceleration (\ref{acc}) can also be derived from any arc length of the circle in time $\Delta t$ via
\be
v=\frac{\theta r}{\Delta t},\qquad\quad
a=\frac{\theta v}{\Delta t}.
\ee
This is similar to methods used in most textbooks, by taking the limit $\Delta t\rightarrow 0$. 
However, the generalization of Newton's {\it Waste Book} construction\cite{erl94} shows that the acceleration (\ref{acc}) 
would remain the same if one replaces the arc length
by its corresponding {\it chord}. This is shown in Fig.1. 

\begin{figure}
	\centerline{\includegraphics[width=0.7\linewidth]{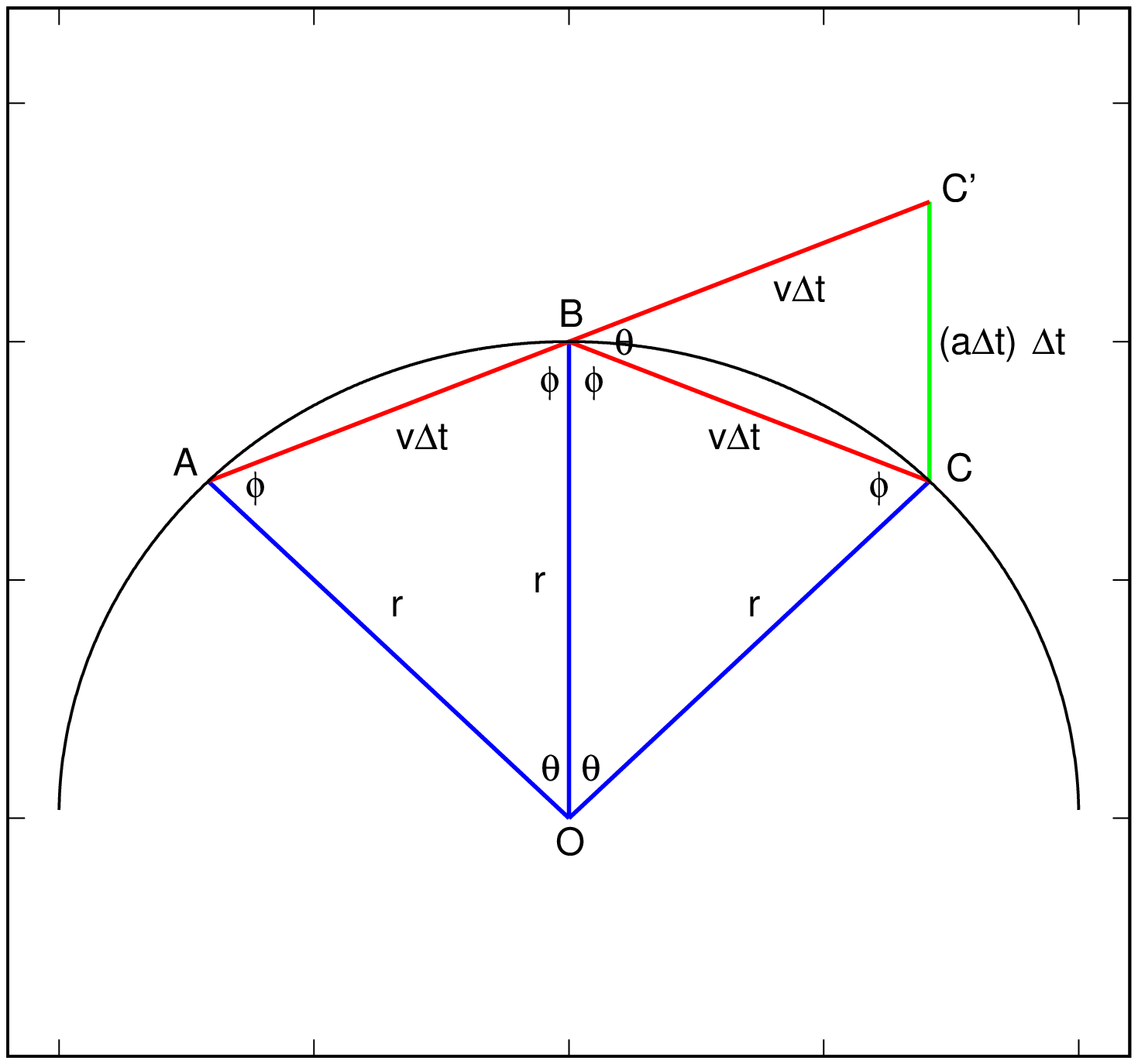}}
	\caption{
		Newton's polygonal orbit with triangle $ABO$ similar to $C'CB$ because $\theta+2\phi=180^\circ$.
		\label{oned}}
\end{figure}

In Fig.1, one seeks to find the acceleration {\bf a} such that
the change in velocity $\Delta{\bf v}={\bf a}\Delta t$ will produce
a polygonal orbit of uniform speed $v$ on a circle of radius $r$.
Starting at $A$, in time $\Delta t$ and uniform speed $v$, the particle moves to $B$ along the
chord $AB=v\Delta t$ with center angle $\theta$. If there were no force acting on it, it would
move to $C'$ in the next time interval unimpeded. 
However, if it were to land on $C$, also on the circle, 
of equal chord length $v\Delta t$ from
$B$, then there must be a change of velocity ${\bf a}\Delta t$ {\it at} $B$, toward the center,
resulting in a change of displacement
from $C'$ to $C$ of length $(a\Delta t)\Delta t$. 
Since $AB=v\Delta t$ is the base of an 
isosceles triangle $AOB$ with two side lengths $r$, 
and $C'C=(a\Delta t)\Delta t$ is the base of a similar isosceles triangle $C'BC$ with
two side lengths $v\Delta t$, both must have the same base to side ratio
\be
\frac{a(\dt)^2}{v\dt}=\frac{v\dt}{r},
\la{om}
\ee
thereby yielding (\ref{acc}) {\it independent} of $\dt$. Thus (\ref{acc}) is true for
any regular $N$-side polygon on the circle. 
Since the length of the polygon is $Nv\dt$, in the limit of $N\rightarrow\infty$
and $\dt\rightarrow 0$ such that $N\dt=T$, the polygon becomes a circle with circumference $2\pi r$ 
and therefore $v T=2\pi r$. It would seem that this should be the end of Newton's derivation.

Surprisingly, according to historian Richard Westfall,\cite{wes71} Newton's did not stop at (\ref{om}).
Instead, Newton multiplies it by $N$, getting $aN\dt/v=vN\dt/r$ and arrives at the velocity circle
$aT/v=vT/r=2\pi$! Westfall then concludes that when this last result is divided by $T=2\pi r/v$, one
would get $a=v^2/r$!! Erlichson\cite{erl91} disputed whether Newton actually did such a division, but
it is indisputable that Newton had known about the velocity circle $aT/v=2\pi$ since the
{\it Waste Book} time around/after 1664.

\begin{figure}
	\centerline{\includegraphics[width=0.7\linewidth]{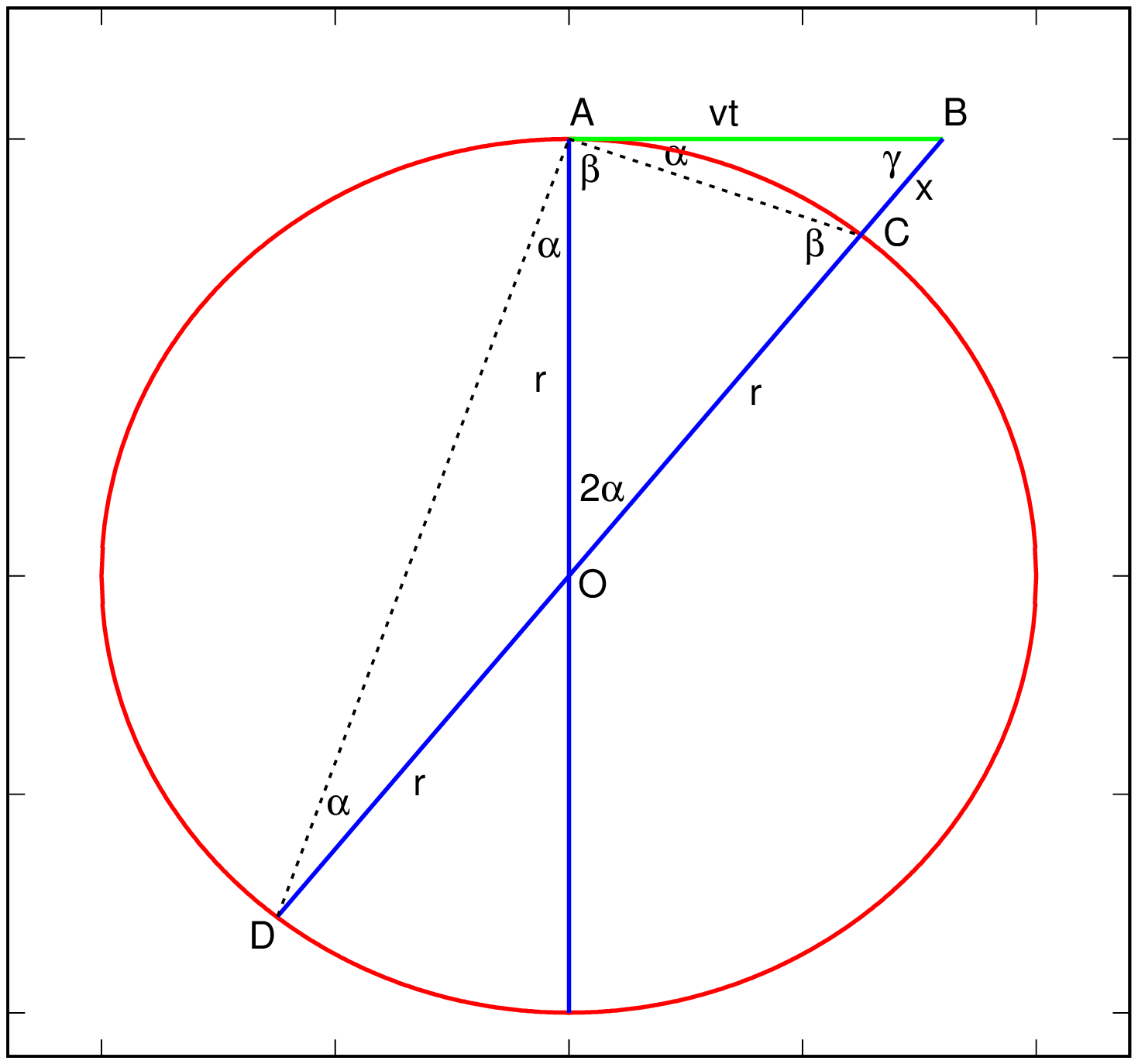}}
	\caption{
		Huygens' key diagram with triangle $ABC$ similar to $ABD$ because
		both have the same two angles $\gamma$ and $\alpha$.
		\label{huy}}
\end{figure}

The other simple derivation, also independently found by many\cite{jon69,pat75,cor12}
(including Tipler's text\cite{tip91}),
was first given geometrically by Huygens\cite{erl94,huy,sti01}, also more than 350 years ago. 
Imagine a stone tied by a tiny string to the rim of
a merry-go-around of radius $r$ rotating at constant rim speed $v$, as shown in Fig.\ref{huy}.
If the string were cut, the stone would fly off the tangent at constant speed $v$ to $B$,
a distance of $vt$ in a {\it short} time $t$. For such a short time, the merry-go-around 
would have rotated from $A$ to $C$ and would view the stone as ``fleeing'' away from the center 
to a distance $x$ from the rim given by
\ba
(r+x)^2&=&(vt)^2+r^2\nn\\
(x+2r)x&=&(vt)^2.
\la{xr}
\ea
Neglecting $x$ relative to $2r$, this center-fleeing distance is
\ba
x &=&\frac12 (\frac{v^2}{r})t^2.
\la{centx}
\ea 
Instead of algebra, Huygens cleverly observes that $ABC$ and $ABD$
are similar triangles and therefore immediately arrive at (neglecting $x$ in $BD=2r+x$)
\ba
\frac{BC}{AB}=\frac{AB}{BD}\quad\rightarrow\quad x=\frac{(vt)^2}{2r}.
\ea
Huygens view this center-fleeing distance $x$ as due to a constant {\it centrifugal} acceleration $v^2/r$.
This explanation from the rotating frame is clear and obvious. (It is not clear why Ref.\onlinecite{hen00} would state that Huygens' method is ``not elegant" yet praises the {\it identical} algebraic expression 
in Tipler's text as ``presented elegantly".) 
On the other hand, if one
stays in the inertial frame, then $x$ is the distance {\it prevented} from happening by a centripetal acceleration
so that the stone would stay put on the rim. This is a less obvious
point-of-view and is the fundamental reason why students find centripetal acceleration more difficult to understand.

\begin{figure}
	\centerline{\includegraphics[width=0.7\linewidth]{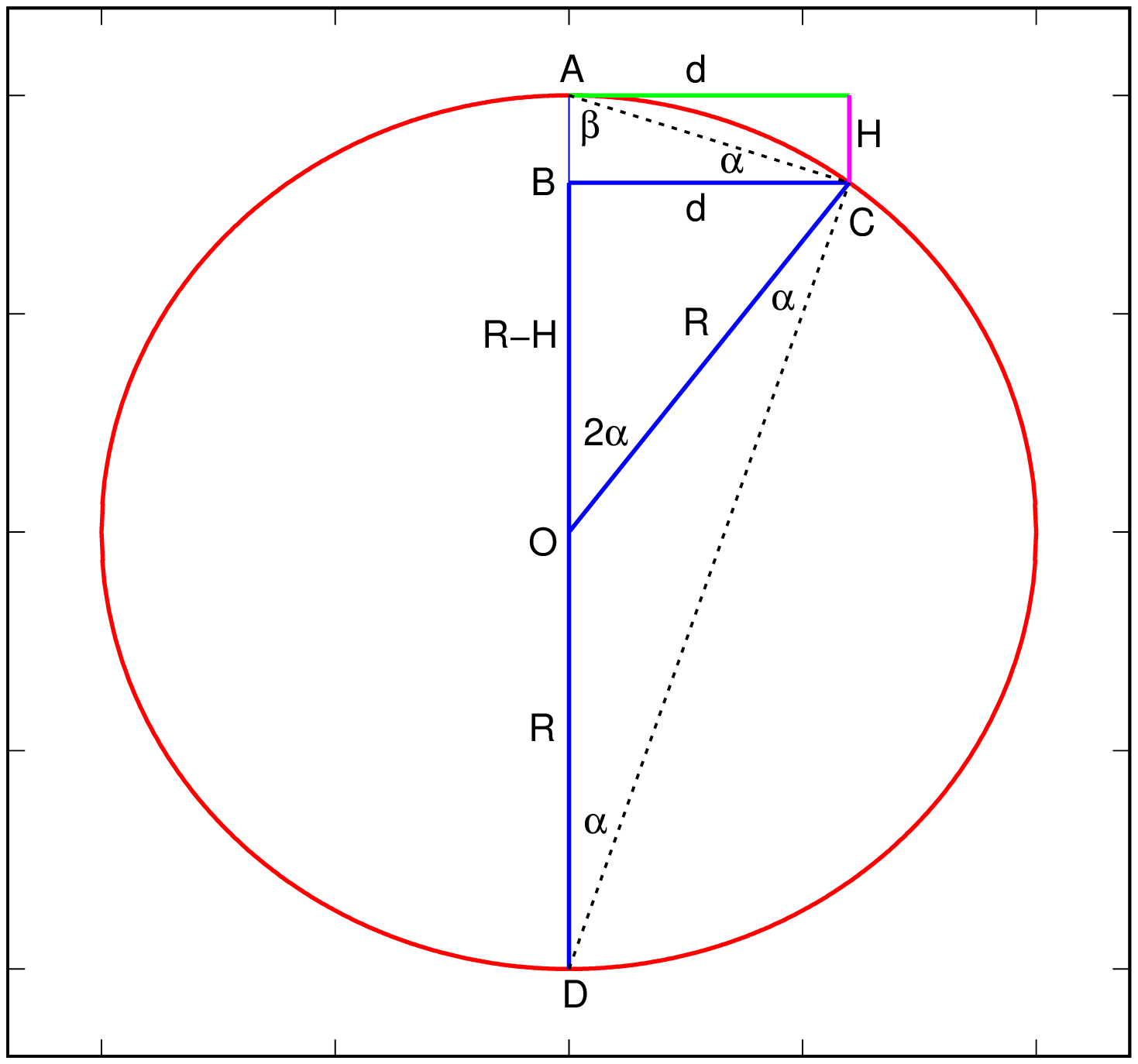}}
	\caption{
	The dropping distance $H$ due to earth's curvature. Triangle $ABC$ is similar to
	$CBD$ because both are right triangles and $\alpha+\beta=90^\circ$.
		\label{earth}}
\end{figure}

After reviewing the above two historical derivations, this work proposes the following 
more physical picture. From projectile motion, one learns that a baseball 
thrown horizontally with velocity $v$ would travel a horizontal distance $d$ in time $t=d/v$.
During this time, the ball would free fall a height of 
\be
h=\frac12 g t^2=\frac12 g\frac{d^2}{v^2}.
\la{fh}
\ee
If $h$ equals to the the initial height of the thrown ball, then the ball hits the ground
at a distance $d$ from its original location. This is elementary to all students.
However, the earth is not flat. As shown in Fig.\ref{earth}, over a horizontal distance $d$
earth's ``ground'' drops below the horizon by a distance $H$.
This $H$ can be solved from the obvious right triangle $BCO$. However, it is also easy to
follow Huygens, by noting that triangle $ABC$ is similar to triangle $CBD$, and therefore
\ba
\frac{AB}{BC}=\frac{BC}{BD}\quad\rightarrow\quad H=\frac{d^2}{2R}.
\ea
If a baseball were thrown with velocity $v$ such that the dropping of the ball $h$ exactly matches 
the dropping of the earth $H$, then the ball will never hit the ground. This velocity is
\be
v=\sqrt{gR}.
\la{orbv}
\ee
At this velocity, the baseball becomes airborne and orbits the earth in a circular
orbit with radius $R$, always free falling toward the center of earth with acceleration $g$.
Generalizing $g$ to any center-seeking acceleration $a$ and earth's radius $R$ to any radius $r$,
then (\ref{orbv}) reproduces the needed centripetal acceleration (\ref{acc}).
In this derivation, the orbit is circular because the object is constantly falling
toward the center, but just keep on missing the ``ground''.

\end{document}